\documentclass[traditabstract]{aa}

\usepackage{graphicx,float}
\usepackage{tabularx}
\usepackage{latexsym,amsmath,amssymb}
\usepackage{natbib}
\usepackage{verbatim}

\bibpunct{(}{)}{;}{a}{}{,}

\def \george    {G. Younes}
\def \laurence  {L. Boirin}
\def \bassem    {B. Sabra}

\def \strasbourg {Observatoire Astronomique de Strasbourg, 11 rue de
l'Universit\'e, F-67000 Strasbourg, France}

\def \ndu {Department of Sciences,  Notre Dame University-Louaize, P.O.Box 72, Zouk Mikael, Lebanon}

\newcommand {\xmm} {XMM-Newton}

\def\ergscm{\hbox{erg s$^{-1}$ cm}}

\def\degmark{$^\circ$}
\def \rsun {\ifmmode$R$_{\odot}\else R$_{\odot}$}

\def \hcm {\hbox {\ifmmode $ atoms cm$^{-2}\else atoms cm$^{-2}$\fi}}

\def\approxgt{\mathrel{\hbox{\rlap{\lower.55ex \hbox {$\sim$}}
        \kern-.3em \raise.4ex \hbox{$>$}}}}
\def\approxlt{\mathrel{\hbox{\rlap{\lower.55ex \hbox {$\sim$}}
        \kern-.3em \raise.4ex \hbox{$<$}}}}

\newcommand {\eg} {{\it e.g.}}

\def\arcmin{\hbox{$^\prime$}}
\def\arcsec{\hbox{$^{\prime\prime}$}}
\newcommand {\ergs} {erg~s$^{-1}$}
\newcommand {\ergcms} {erg cm$^{-2}$ s$^{-1}$}
\newcommand {\chisq} {$\chi ^{2}$}

\newcommand {\fetfive} {\ion{Fe}{xxv}}
\newcommand {\fetsix} {\ion{Fe}{xxvi}}

\def \nineteen {XB\,1916$-$053}
\def \mxb {MXB\,1658$-$298}
\def \bigdip {X\,1624$-$49}
\def \twelve {X\,1254$-$690}

\def \exo {EXO\,0748$-$676}

\def \thirteen {4U\,1323$-$62}

\def \useventeen {4U\,1746$-$371}

\def \src {XTE~J1710$-$281}

\begin{document}

\title{An XMM-Newton view of the dipping low-mass X-ray binary \src}

\author{\george\inst{1} \and \laurence\inst{1} \and \bassem\inst{2}}

\institute{\strasbourg \and \ndu}

\date{Received / Accepted}

\authorrunning{G. Younes et al.}

\titlerunning{XMM-Newton observation of \src}


\abstract{Studying the spectral changes during the dips exhibited by
  almost edge-on, low-mass X-ray binaries (LMXBs) is a powerful means
  of probing the structure of accretion disks.  The \xmm, Chandra, or
  Suzaku discovery of absorption lines from \fetfive\ and other
  highly-ionized species in many dippers has revealed a highly-ionized
  atmosphere above the disk.  A highly (but less strongly) ionized
  plasma is also present in the vertical structure causing the dips,
  together with neutral material. We aim to investigate the spectral
  changes during the dips of \src, a still poorly studied LMXB known
  to exhibit bursts, dips, and eclipses. We analyze the archived
  XMM-Newton observation of \src\ performed in 2004 that covered one
  orbital period of the system (3.8~hr). We modeled the spectral
  changes between persistent and dips in the framework of the partial
  covering model and the ionized absorber approach. The persistent
  spectrum can be fit by a power law with a photon index of
  $1.94\pm0.02$ affected by absorption from cool material with a
  hydrogen column density of
  $(0.401\pm0.007)\times10^{22}$~cm$^{-2}$. The spectral changes from
  persistent to deep-dipping intervals are consistent with the partial
  covering of the power-law emission. Twenty-six percent of the
  continuum is covered during shallow dipping, and 78\%\ during deep
  dipping. The column density decreases from $77_{-38}^{+67}\times
  10^{22}$~cm$^{-2}$ during shallow dipping to $(14 \pm 2)\times
  10^{22}$~cm$^{-2}$ during the deep-dipping interval. We do not
  detect any absorption line from highly ionized species such as
  \fetfive. However, the upper-limits we derive on their equivalent
  width (EW) are not constraining. Despite not detecting any narrow
  spectral signatures of a warm absorber, we show that the spectral
  changes are consistent with an increase in column density and a
  decrease in ionization state of a highly-ionized absorber,
  associated with an increase in column density of a neutral absorber,
  in agreement with the recent results found in other dippers.  In
  \src, the column density of the ionized absorber increases from
  $4.3_{-0.5}^{+0.4}\times 10^{22}$~cm$^{-2}$ during shallow dipping
  to $11.6_{-0.6}^{+0.4}\times 10^{22}$~cm$^{-2}$ during deep dipping,
  while the ionization parameter decreases from $10^{2.52}$ to
  $10^{2.29}$~\ergscm. The parameters of the ionized absorber are not
  constrained during persistent emission.  The neutral absorber only
  slightly increases from $(0.410 \pm 0.007)\times 10^{22}$ ~cm$^{-2}$
  during persistent emission to $(0.420 \pm 0.009)\times 10^{22}$
  ~cm$^{-2}$ during shallow dipping and to $(0.45 \pm 0.03)\times
  10^{22}$ ~cm$^{-2}$ during deep dipping. The warm absorber model
  better accounts for the $\sim1$~keV depression visible in the pn
  dipping spectra, and naturally explains it as a blend of lines and
  edges unresolved by pn. A deeper observation of \src\ would enable
  this interpretation to be confirmed.}

\keywords{Accretion, accretion disks -- Stars: individual: \src\ -- X-rays: general}

\maketitle


\section{Introduction}
\label{sec:intro}

\src~was discovered in 1998 serendipitously by RXTE at a position
consistent with the location of the unidentified \textit{ROSAT} source
1RXS J171012.3-280754 \citep{1710:markwardt98iau}.  Its flux varies
between about 2 and 10~mCrab on timescales of about 30~days
\citep{1710:markwardt01conf}. \src~shows X-ray bursts indicating that
the compact object is a neutron star and that the system could lie at
a distance of 15--20~kpc \citep{1710:markwardt01conf} or 12--16~kpc
\citep{galloway06apjs}.  Its light curve presents eclipses lasting
410~s and dipping activity, both recurring at the orbital period of
3.28~hr. The dips are believed to be due to obscuration in the
thickened, azimuthally structered, outer regions of an accretion disk
\citep{1916:white82apjl}. The presence of eclipses indicates that the
source is viewed close to edge-on, at an inclination angle, $i$, of
75\degmark--80\degmark \citep{frank87aa}, where $i$ is defined as the
angle between the line of sight and the rotation axis of the accretion
disk.  The RXTE PCA spectrum is consistent with either thermal
bremsstrahlung (kT = 14 $\pm$ 3 keV) or a power law (photon index 1.8
$\pm$ 0.1), with interstellar absorption $N_{H} < 2\times10^{22}$
~cm$^{-2}$ \citep{1710:markwardt98iau}.

The 1--10~keV spectra of most of the dip sources become harder during
dipping. However, these changes are inconsistent with a simple
increase in photo-electric absorption by cool material. These changes
have been often modeled by the ``progressive covering'', or ``complex
continuum'' approach \citep[e.g.,][]{1916:church97apj}, in which the
X-ray emission is assumed to originate from a point-like blackbody, or
disk-blackbody component, and from a power-law component accounting
for an extended corona.  This approach models the spectral changes
during dipping intervals by the partial and progressive covering of
the extended component by an opaque neutral absorber.  The absorption
of the point-like component is allowed to vary independently from that
of the extended component.

The improved sensitivity and spectral resolution of {\it Chandra},
XMM-Newton and Suzaku is allowing narrow absorption features from
highly ionized Fe and other metals to be observed from a growing
number of X-ray binaries.  In particular, \fetfive\ (He-like) or
\fetsix\ (H-like) 1s-2p resonant absorption lines near 7~keV were
reported from several micro-quasars and neutron star LMXBs
\citep[e.g.,][]{1630:kubota07pasj,1658:sidoli01aa} that are almost all
viewed close to edge-on, many of them being dippers \citep[see
e.g. Table~5 of ][]{1916:boirin04aa}.  This indicates that the highly
ionized plasma probably originates in an accretion disk atmosphere or
wind, which could be a common feature of accreting binaries but
primarily detected in systems viewed close to edge-on.  During
the dips, absorption lines are also detected and correspond to
electronic transitions from less ionized species than during the
persistent intervals. \citet{1323:boirin05aa} demonstrated that an
increase in column density and a decrease in the ionization state of a
highly-ionized absorber, associated with the increase in column
density of a local neutral absorber could model the changes between
persistent and dipping intervals both in the X-ray continuum and the
narrow absorption features of \thirteen. This result was successfully
applied to \twelve, \bigdip, \mxb, \useventeen\ and \nineteen~observed
with \xmm\ \citep{diaztrigo06aa}.

Here, we report on a spectral study of the archived \xmm\ observation
of the dipping LMXB \src.  We report spectral hardening from
persistent to dipping intervals and model it in the framework of the
partial covering appraoch and of the warm absorber approach.


\section{Observation and data reduction}
\label{sec:reduction}


\subsection{Data reduction}

The XMM-Newton Observatory \citep{jansen01aa} includes three
1500~cm$^2$ X-ray telescopes each with an European Photon Imaging
Camera (EPIC) at the focus.  Two of the EPIC imaging spectrometers use
MOS CCDs \citep{turner01aa} and one uses pn CCDs \citep{struder01aa}.
Reflection Grating Spectrometers \citep[RGS,][]{denherder01aa} are
located behind two of the telescopes while the 30--cm optical monitor
(OM) instrument has its own optical/UV telescope\citep{mason01aa}.
\src\ was observed by XMM-Newton for 13~ks on 2004 February 22 between
09:55 and 13:16~UTC. The optical monitor instrument was operated in
imaging mode with the U and UVW1 filters (PSF FWHM of 1.55 and
2.0\arcsec, respectively). The U filter bandpass is between $\sim$300
and 400 nm while the UVW1 filter covers the $\sim$230--370 nm
wavelength range. The EPIC-pn and MOS cameras were operated in full
frame mode. We concentrate here on the analysis of the EPIC--pn data
since they are not affected by pile-up, contrary to MOS1 and MOS2 data
where the total maximum count rate to avoid deteriorated response due
to photon pile-up is 150 $s^{-1}$, lot less than the total maximum
count rate in the pn case, 1000 $s^{-1}$. We also use RGS data. All
data products were obtained from the XMM-Newton public archive and
reduced using the Science Analysis System (SAS) version 8.0. Few
intervals of enhanced solar activity were present. However, we did not
discard any of them since the background count rate represents at
maximum 3\% of the source persistent rate.

We selected only single and double events (patterns 0 to 4) from the
pn data. We extracted source events from a circle of $40\arcsec$
radius centered on the source. Background events were obtained from a
circle of the same radius on the same CCD as the source but centered
away. We generated response matrix files using the SAS task
\textit{rmfgen}, while ancillary response files were generated using
the SAS task \textit{arfgen}. We rebinned the EPIC--pn spectra to
oversample the full-width at half maximum (FWHM) of the energy
resolution by a factor 3, and to have a minimum of 25 counts per bin
to allow the use of the $\chi^2$ statistic.

The SAS task \textit{rgsproc} was used to produce calibrated RGS event
lists, spectra, and response matrices. RGS order 2 is not used in the
analysis because it shows few source events detection. The RGS 1st
order spectrum does not show evidence for any narrow spectral features
therefore it was mainly used to check for consistency with the EPIC-pn
spectra and to better constrain the low-energy part of the
spectrum. It was rebinned to have a minimum of 25 counts per bin.


\subsection{Spectral analysis}
\label{sec:spectralanalysis}

We use EPIC-pn spectra in the energy range 0.2--10~keV. We restrict
the RGS spectra (RGS1 and RGS2, order~1) to the energy range of
0.5--2.0~keV because very few source events are detected below
0.5~keV.

Spectral analysis was performed using XSPEC \citep{arnaud96conf}
version 12.0. The updated photo-electric cross sections and the
revised solar abundances of \citet{wilms00ApJ} (``abund wilm'' command
within XSPEC) are used throughout to account for absorption by neutral
gas. Spectral uncertainties are given using $\Delta$\chisq\ of 2.71,
corresponding to 90\% confidence for one interesting parameter, and to
95\% confidence for upper limits. All EWs are quoted with positive
values for absorption features.


\section{Results}


\subsection{Source position}

The second XMM-Newton serendipitous source catalog (2XMM)
\citep{watson08:2xmmi} gives \src\ a position of
R.A.~=~17h10\arcmin12.532\arcsec,
Decl.~=~-28\degmark07\arcmin50.95\arcsec\ (equinox 2000.0) with a 1
sigma uncertainty of 1\arcsec\ in both R.A. and Decl. There is no
source consistent with this X-ray position in the infra--red 2MASS
catalog, neither in the optical image derived from the XMM-Newton
optical monitor, nor in the USNO catalog which is the deepest optical
catalog that we cross--correlated with the 2XMM using the software
XCAT--DB \citep{pineau08aspc}.


\subsection{Lightcurve and hardness ratio}

\begin{figure*}[!ht]
\centerline{\includegraphics[angle=0,width=1\textwidth]{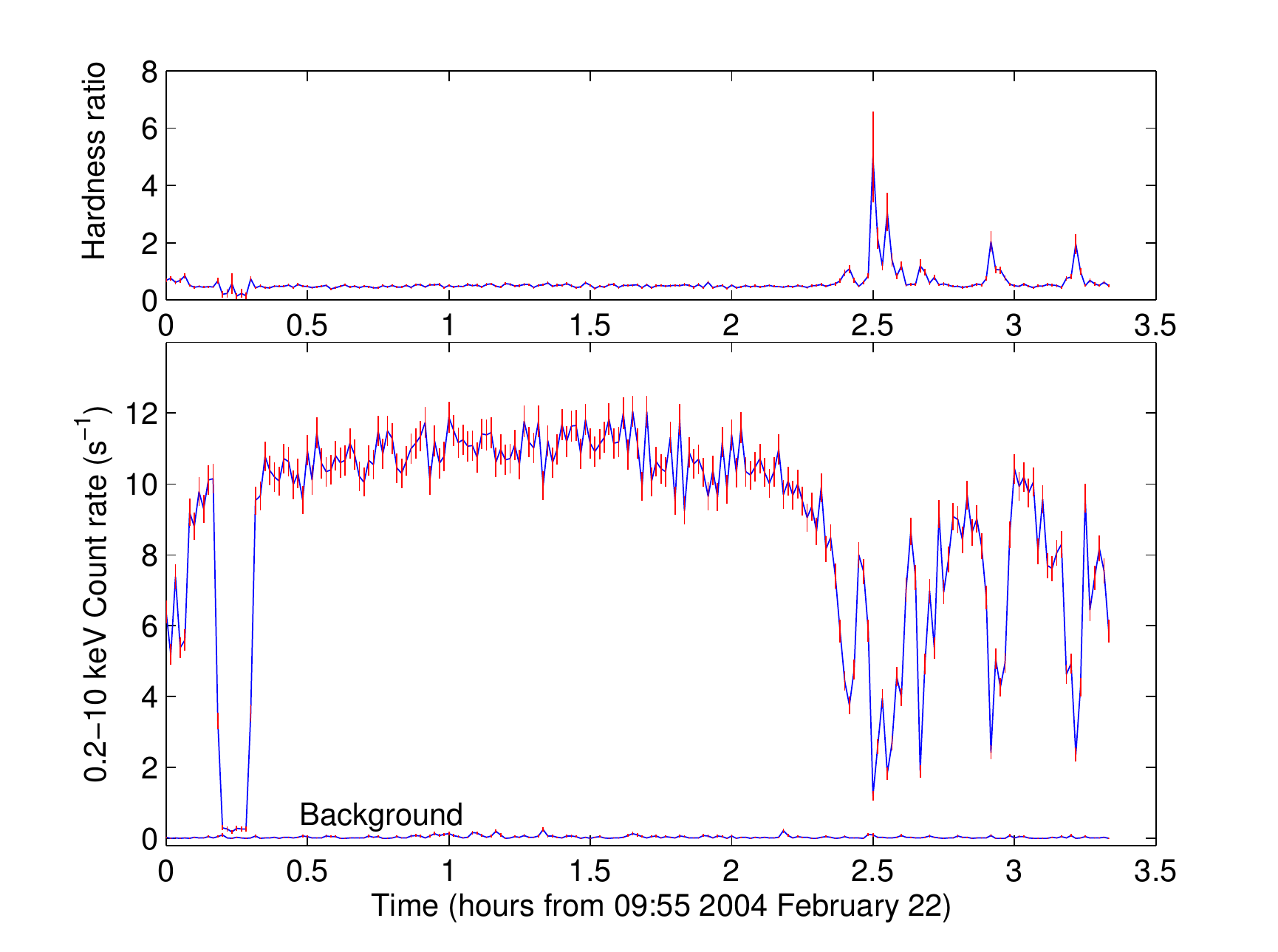}}
\caption{The lower panel shows the 0.2--10~keV EPIC--pn light curve of
\src.  An eclipse and the dipping activity are clearly visible. The
0.2--10~keV background light curve is also displayed.  The upper panel
shows the hardness ratio (counts in the 2.0--10~keV band divided by
those between 0.2--2.0~keV).  The dips are associated with hardening.
The binning is 60~s in each panel.}
\label{LC-XTEJ1710}
\end{figure*}

The EPIC--pn 0.2--10 keV source and background light curves are shown
in the lower panel of Fig. \ref{LC-XTEJ1710} with a binning of 60
s. The observation covers almost one orbital period (3.28 hr) of the
system. One eclipse lasting for 410~s is visible, as well as some
dipping activity at the biginning and the end of the observation. The
interval where the count rate remains approximately constant is
referred to as the persistent emission.

We define the hardness ratio as the counts in the 2.0--10 keV energy
range (the ``hard'' energy band), divided by the counts in the
0.2--2.0 keV energy range (the ``soft'' energy band).  The hardness
ratio evolution as a function of time is shown in the upper panel of
Fig. \ref{LC-XTEJ1710}. An increase in hardness during dipping is
observed. This indicates that the number of counts in the soft band
has decreased more than that in the hard band. This suggests that
absorption is the main ingredient of the dipping activity. The deepest
troughs of the dips are clearly associated with the strongest
hardening, and hence the strongest absorption.  

Based on the entire lightcurve, we extracted the spectrum of the
deep-dipping state from time intervals where the count rate is less
than 5~counts~s$^{-1}$. We extracted the shallow-dipping spectrum from
time intervals with count rates between 5 and
10~counts~s$^{-1}$. Finally, we extracted the spectrum of the
persistent state from time intervals with count rates greater than
10~counts~s$^{-1}$. Of course, the eclipse interval is when the count
rate approached that of the background for a total of 410~s.


\subsection{Persistent spectrum}
\label{perspecsec}

\begin{figure}[!th]
\centerline{\includegraphics[angle=0,width=0.5\textwidth]{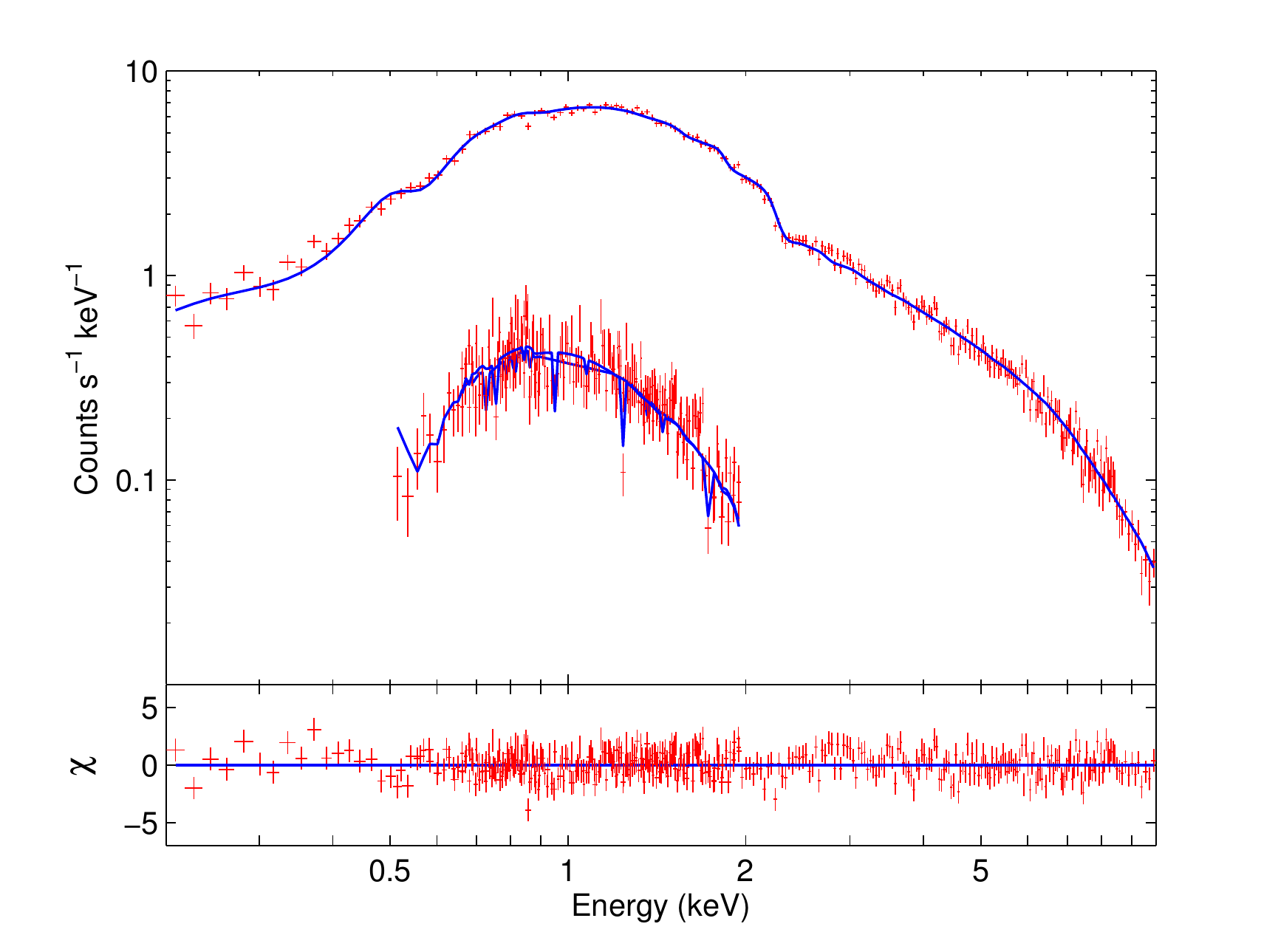}}
\caption{The upper panel shows the EPIC pn spectrum (top), the RGS1
and RGS2 first order spectra (bottom) of \src\ during persistent
intervals. The solid line is the best-fit absorbed power-law
model. The lower panel shows the spectral residuals from the best-fit
model in terms of sigmas.}
\label{rgspnperspectra}
\end{figure}

We fit simultaneously the pn persistent spectrum, the RGS1 first order
spectrum and the RGS2 first order spectrum. We include a
multiplicative constant factor, fixed to 1 for the EPIC-pn spectrum
but allowed to vary for each RGS spectrum, to take account for cross
calibration uncertainties. An absorbed blackbody model or an absorbed
disk-blackbody did not give acceptable fits to the persistent spectrum
(reduced $\chi^{2}$ of: 14 for 430 degrees of freedom (d.o.f.) and 5
for 430 d.o.f. respectively).  An absorbed power law
(\textit{tbabs*pow} within XSPEC) was able to fit the spectra well
with a reduced $\chi^{2}$ of 1.17 for 430 d.o.f.
(Fig. \ref{rgspnperspectra}). A model consisting in two emission
components, a power law and a blackbody, modified by a neutral
absorber gave a reduced $\chi^{2}$ of 1.16 for 428 d.o.f. An F-test
indicates that the probability for such an improvement (compared to
the fit with an absorbed power law) to occur by chance is
0.16. Therefore, the blackbody component is not significant and was
not kept in the model. The best fit photon index is $1.94\pm0.02$
while the neutral hydrogen column density is $(0.401\pm0.007)\times
10^{22}$~cm$^{-2}$. We refer hereafter to this absorption as
foreground absorption, it includes contributions from the interstellar
medium and/or absorption intrinsic to the system itself. The
normalization of the power law is
$(1.21~\pm~0.02)~\times~10^{-2}$~photons
~keV~$^{-1}$~cm$^{-2}$~s$^{-1}$ and the constant factors are
$1.11\pm0.05$ and $1.10\pm0.04$ for the RGS1 and RGS2 data
respectively.

In the 0.2--10 keV energy range, we derive an absorbed flux of
$4.69\times10^{-11}$~\ergcms, an unabsorbed flux of
$7.79\times10^{-11}$~\ergcms, and a luminosity of $L_{x} ~=
~2.4~\times~10^{36}$~\ergs\ assuming a source distance of 16 kpc.

We do not detect absorption lines due to highly ionized elements
(\fetfive ~or \fetsix ~near 7~keV) in the persistent emission
spectrum.


\subsection{Persistent, shallow and deep-dipping spectra}
\label{noabsorptionEW}

We consider the dipping activity as due to the passage of a thickened
part of the disk in front of the underlying X-ray source, at each
orbital rotation of the system. Since the count rate and the hardness
ratio of \src\ are constant during the persistent interval (see
Fig. \ref{LC-XTEJ1710}), we may assume that the underlying X-ray
emission does not change significantly during the whole observation,
and that the spectral changes observed during the dipping interval are
indeed arising from the passage of some absorbing material in the
line-of-sight. The rapid variability of the spectral changes during
the dipping interval indicates that this effect dominates over a
potentially coincident smooth and slow spectral transition of the
underlying X-ray source itself.

Therefore, to explain the spectral changes during dipping, we fit the
persistent, shallow and deep-dipping spectra simultaneously, and we
link the parameters of the powerlaw component.  For the persistent
interval, we use pn and RGS (order 1) spectra but plot only pn for
clarity. For the dipping intervals, we use only the pn spectra because
too few counts are detected by RGS. In a first step, we allowed the
foreground column density to be different for the three spectra, but
did not obtain any acceptable fit (reduced $\chi^{2}$ of 4.4 for 769
d.o.f.). Consequently, we tested a partial covering approach
(Sect. \ref{parcovsec}) and a warm absorber approach
(Sect. \ref{warmabssec}).

We do not detect any absorption line from highly ionized species such
as \fetfive\ in \src, neither in the persistent spectrum nor in the
dipping spectra. By including a Gaussian profile with a width fixed to
0 at 6.65~keV, the theoretical energy of the \fetfive\ 1s-2p
transition, we derived upper-limits of 114, 50, and 73~eV on the EW of
this line in the persistent, shallow-dipping, and deep-dipping spectra,
respectively. Since lines from \fetfive\ or other species were
detected with an EW down to ~25~eV in other binaries \citep[\eg\
][]{1323:boirin05aa}, the upper-limits derived for \src\ are not
constraining. The non detection of absorption lines in \src\ could be
explained by a lack of statistics in the current XMM-Newton
observation.


\subsubsection{Progressive covering model}
\label{parcovsec}

\begin{figure*}[!ht]
\centerline{\includegraphics[angle=0,width=1\textwidth]{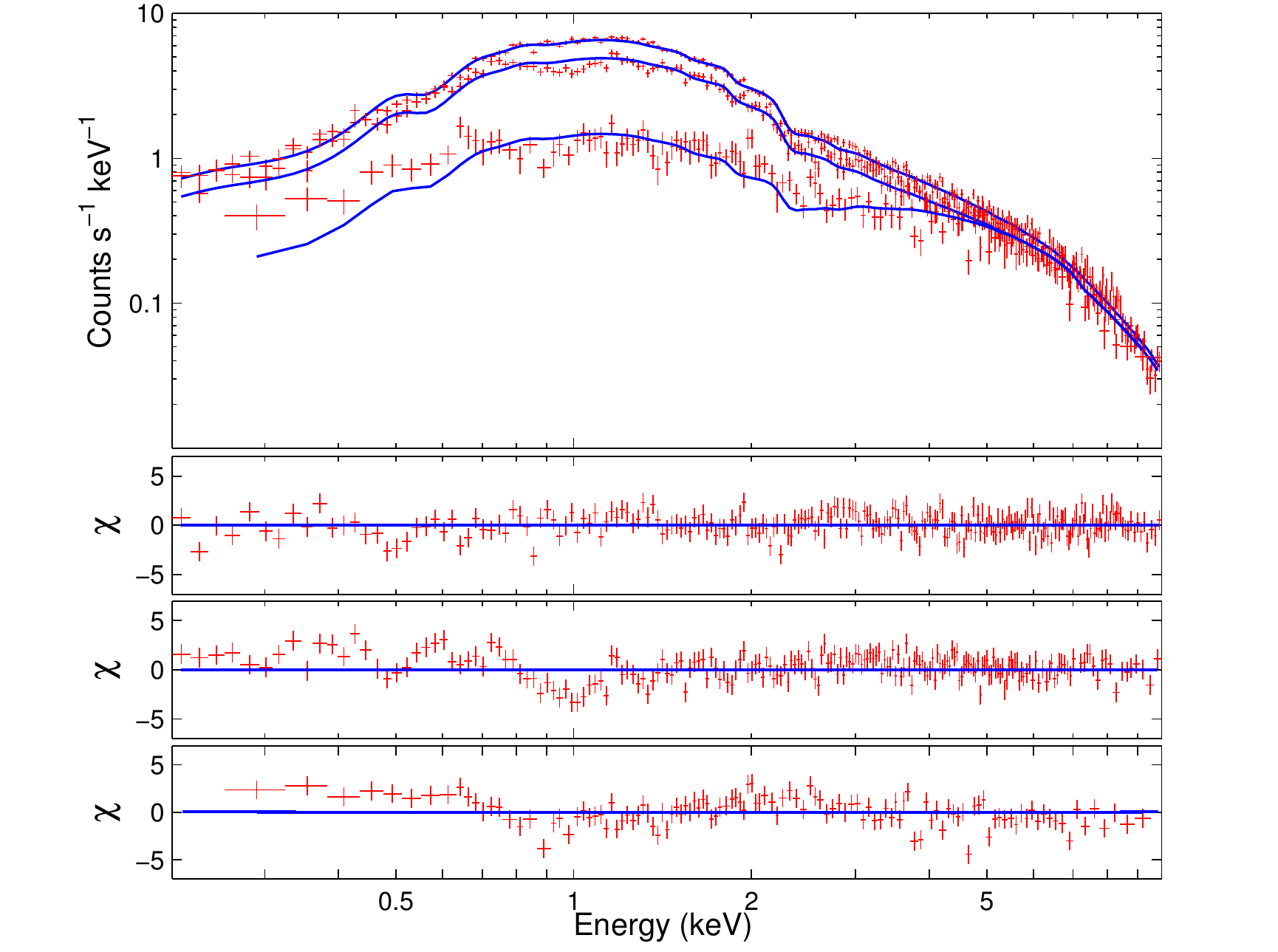}}
\caption{The upper panel shows, from top to bottom, the EPIC pn
persistent, shallow-dipping and deep-dipping spectra of \src, fit
simultaneously with the partial covering model (see
Sect. \ref{parcovsec}). The three other panels show the residuals of
each spectrum from the model, in terms of sigmas.}
\label{partialcoverage}
\end{figure*}

\begin{table*}[!th]
\label{PCtable}
\caption{Best fit spectral parameters for the persistent, the shallow-dipping, and the deep-dipping spectra fit simultaneously with the partial covering model.}
\newcommand\T{\rule{0pt}{2.6ex}}
\newcommand\B{\rule[-1.2ex]{0pt}{0pt}}
\begin{center}{
\begin{tabular}{l c c c c c c}
\hline
\hline
\multicolumn{2}{l}{\textbf{Parameter}} \T \B & \textbf{Per. emission} & & \textbf{Shallow dip} & & \textbf{Deep dip} \\
\hline
\multicolumn{2}{l}{$N_{H}^{fore} ~(10^{22}$~cm$^{-2})$} \T & & & $0.385 \pm 0.007$ & & \\
\multicolumn{2}{l}{Photon index} \T & & & $1.91 \pm 0.02$ & & \\
\multicolumn{2}{l}{Norm.~ \T (photons ~keV$^{-1}$~cm$^{-2}$~s$^{-1}$)} & & & $0.0117 \pm 0.0002$ & & \\
\multicolumn{2}{l}{$N_{H}^{pow}~ (10^{22}$~cm$^{-2})$} \T & 0.0 Fixed & & $77_{-38}^{+67}$ & & $14 \pm 2$ \\
\multicolumn{2}{r}{Significance} \T \B & & & & $2.7~\sigma$ & \\
\multicolumn{2}{l}{$f$} \T \B & 0.0 Fixed & & $0.26 \pm 0.01$ & & $0.776 \pm 0.009$ \\
\multicolumn{2}{r}{Significance} \T \B & & & & $>10~\sigma$ & \\
\hline
\multicolumn{7}{c}{Reduced $\chi^{2} = 1.55$ for 767 d.o.f.} \T \B \\
\hline
\end{tabular}}
\end{center}
\begin{list}{}{}
\item[Note:]The foreground hydrogen column density, $N_{H}^{fore}$, the photon index and the normalization of the power law are linked between persistent, shallow and deep-dipping emission. The covering fraction $f$ and the hydrogen column density, $N_{H}^{pow}$, are allowed to vary to explain the spectral changes in the spectra. Uncertainties are given at a 90\% confidence level. We also indicate the significance of the change of $N_{H}^{pow}$ and $f$ between shallow and deep-dipping spectra.
\end{list}
\end{table*}

In this approach, we allow the continuum emission to be partially
covered by some neutral material during dipping. Since the continuum
emission of \src\ may be simply represented by a power law, we define
the model of the absorbed spectrum as

\begin{equation}
F(E) = e^{-\sigma(E)N_{H}^{fore}}[I_{pow}(f e^{-\sigma(E)N_{H}^{pow}}+(1-f))]
\end{equation}
 
where $I_{pow}$ is the energy dependent flux of the power-law
component, $N_{H}^{pow}$ is the column density of the absorber
affecting the power law, and $N_{H}^{fore}$ is the column density of
the foreground absorber. The photo-electric absorption cross section,
$\sigma$(E), does not include Thompson scattering.  The covering
fraction, $f$, can vary from 0 to 1 and is unit-less.

The foreground column density, $N_{H}^{fore}$, was linked for the
three emission intervals assuming that the X-ray source is subject to
the same foreground absorption. We fixed the covering fraction, $f$,
and $N_{H}^{pow}$~ to zero during the persistent emission, since this
emission interval is not supposed to undergo any absorption from the
bulge. These two parameters are allowed to vary for the
shallow-dipping and the deep-dipping emission.

The spectra and best fit model are shown in
Fig. \ref{partialcoverage}. Table~ 1 gives the best fit parameter
values. The reduced $\chi^{2}$ of the best fit is 1.55 for 767
d.o.f. The foreground column density is $0.38\times 10^{22}$
~cm$^{-2}$, the photon index is $ 1.91$~ and the normalization of the
power law is $1.17\times
10^{-2}$~photons~keV$^{-1}$~cm$^{-2}$~s$^{-1}$. These values are
consistent with those found for the persistent emission alone
(Sect. \ref{perspecsec}). The covering fraction increases from 26\%~
during the shallow-dipping, to 78\%~ during the deep-dipping interval,
which represents a change at a level of more than 10~$\sigma$.  This
would indicate that the emission region is angularly more extended
than the absorbing region, and that a larger fraction of it is covered
during deep dipping. However, we find that $N_{H}^{pow}$ decreases
from $77_{-38}^{+67}$~cm$^{-2}$ during the shallow dip to $14 \pm
2$~cm$^{-2}$ during the deep dip interval, which represents a change
at a level of 2.7~$\sigma$. This would mean that the absorbing layer
is thiner or less dense during deep dipping than during shallow
dipping.


\subsubsection{Ionized absorber model}
\label{warmabssec}

\begin{figure*}[!ht]
\centerline{\includegraphics[angle=0,width=1\textwidth]{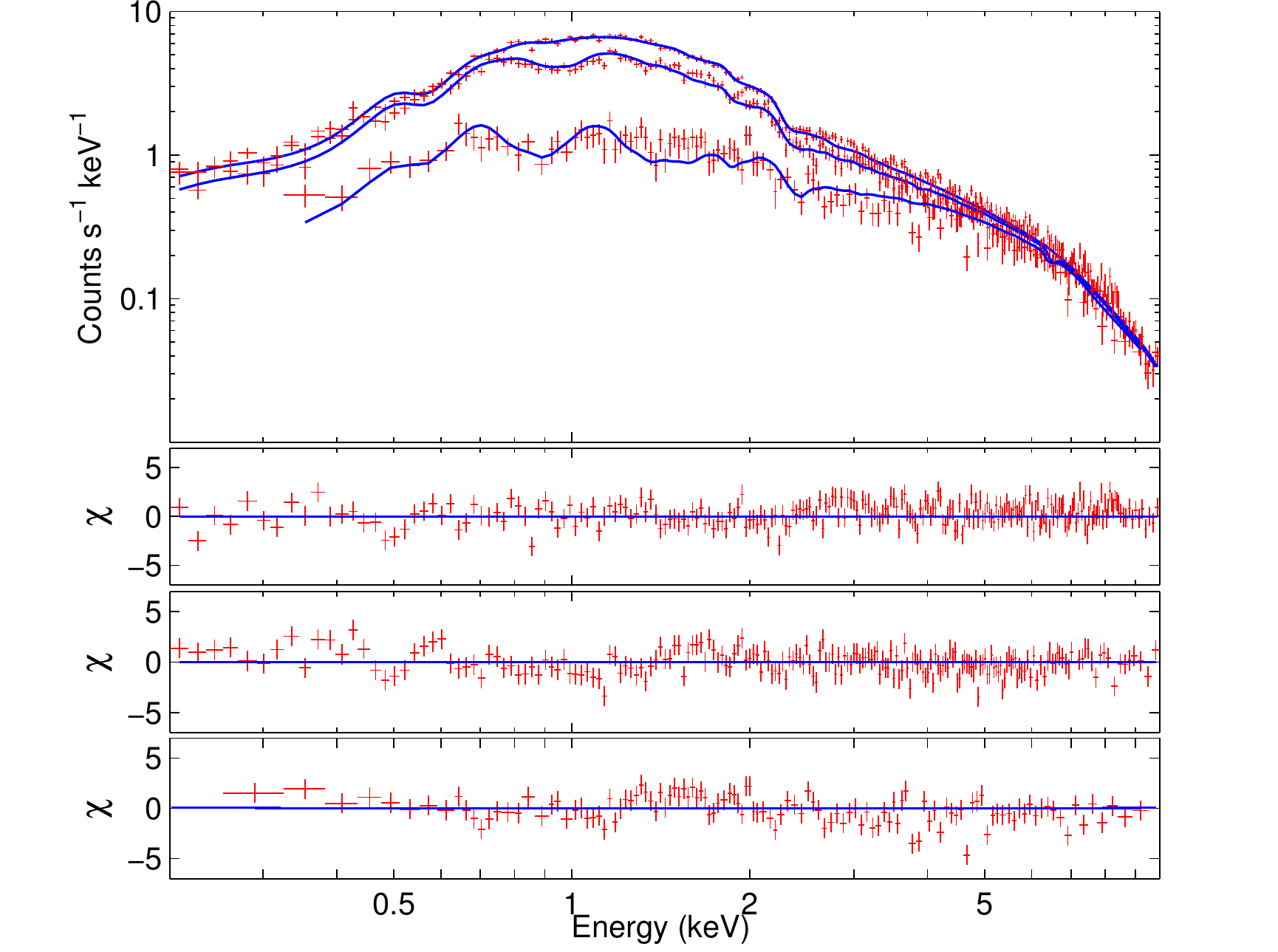}}
\caption{The upper panel shows, from top to bottom, the EPIC pn
persistent, shallow-dipping and deep-dipping spectra of \src, fit
simultaneously with the warm absorber model (see
Sect. \ref{warmabssec}). The three other panels show the residuals of
each spectrum from the model, in terms of sigmas.}
\label{fitwarmabsorber}
\end{figure*}

\begin{table*}[!th]
\label{watable}
\caption{Best fit spectral parameters for the persistent, the shallow-dipping, and the deep-dipping spectra fit simultaneously with the ionized absorber model.}
\newcommand\T{\rule{0pt}{2.6ex}}
\newcommand\B{\rule[-1.2ex]{0pt}{0pt}}
\begin{center}{
\begin{tabular}{l c c c c c c}
\hline
\hline
\multicolumn{2}{l}{\textbf{Parameter}} \T \B & \textbf{Per. emission} & & \textbf{Shallow dip} & & \textbf{Deep dip} \\
\hline
\multicolumn{2}{l}{$N_{H}^{fore}~ (10^{22}$~cm$^{-2})$} \T & $0.410 \pm 0.007$ & & $0.420 \pm 0.009$ & & $0.45 \pm 0.03$ \\
\multicolumn{2}{r}{Significance} \T \B & & $1.4~\sigma$ & & $1.7~\sigma$ & \\
\multicolumn{2}{l}{Photon index} \T & & & $1.98 \pm 0.01$ & & \\
\multicolumn{2}{l}{Norm.~(photons ~keV$^{-1}$~cm$^{-2}$~s$^{-1}$)} \T & & & $0.0124 \pm 0.0002$ & &\\
\multicolumn{2}{l}{$log(\xi)$~(erg ~s$^{-1}$~cm)} \T & 5.0 Frozen & & $2.52_{-0.07}^{+0.05}$ & & $2.29 \pm 0.02$ \\
\multicolumn{2}{r}{Significance} \T \B & & & & $5~\sigma$ & \\
\multicolumn{2}{l}{$N_{H}^{\xi} ~(10^{22}$~cm$^{-2})$} \T \B & 0.32 Frozen & & $4.3_{-0.5}^{+0.4}$ & & $11.6_{-0.6}^{+0.4}$ \\
\multicolumn{2}{r}{Significance} \T \B & & & & $>10~\sigma$ & \\
\hline
\multicolumn{7}{c}{Reduced $\chi^{2} = 1.42$ for 765 d.o.f.} \T \B \\
\hline
\end{tabular}}
\end{center}
\begin{list}{}{}
\item[Note:]The photon index and the normalization of the power law are linked between persistent, shallow and deep dipping. $N_{H}^{fore}$ is the foreground neutral hydrogen column density. $\xi$ is the ionization parameter. $N_{H}^{\xi}$ is the warm absorber hydrogen column density. These parameters are allowed to vary to explain the spectral changes from persistent to deep dipping. Uncertainties are given at a 90\% confidence level. We also indicate the significance of the change of $N_{H}^{fore}$ between the three spectra as well as the significance of the change of $log(\xi)$ and $N_{H}^{\xi}$ between shallow and deep-dipping spectra.
\end{list}
\end{table*}

\begin{figure*}[!ht]
\centerline{\includegraphics[angle=0,width=1\textwidth]{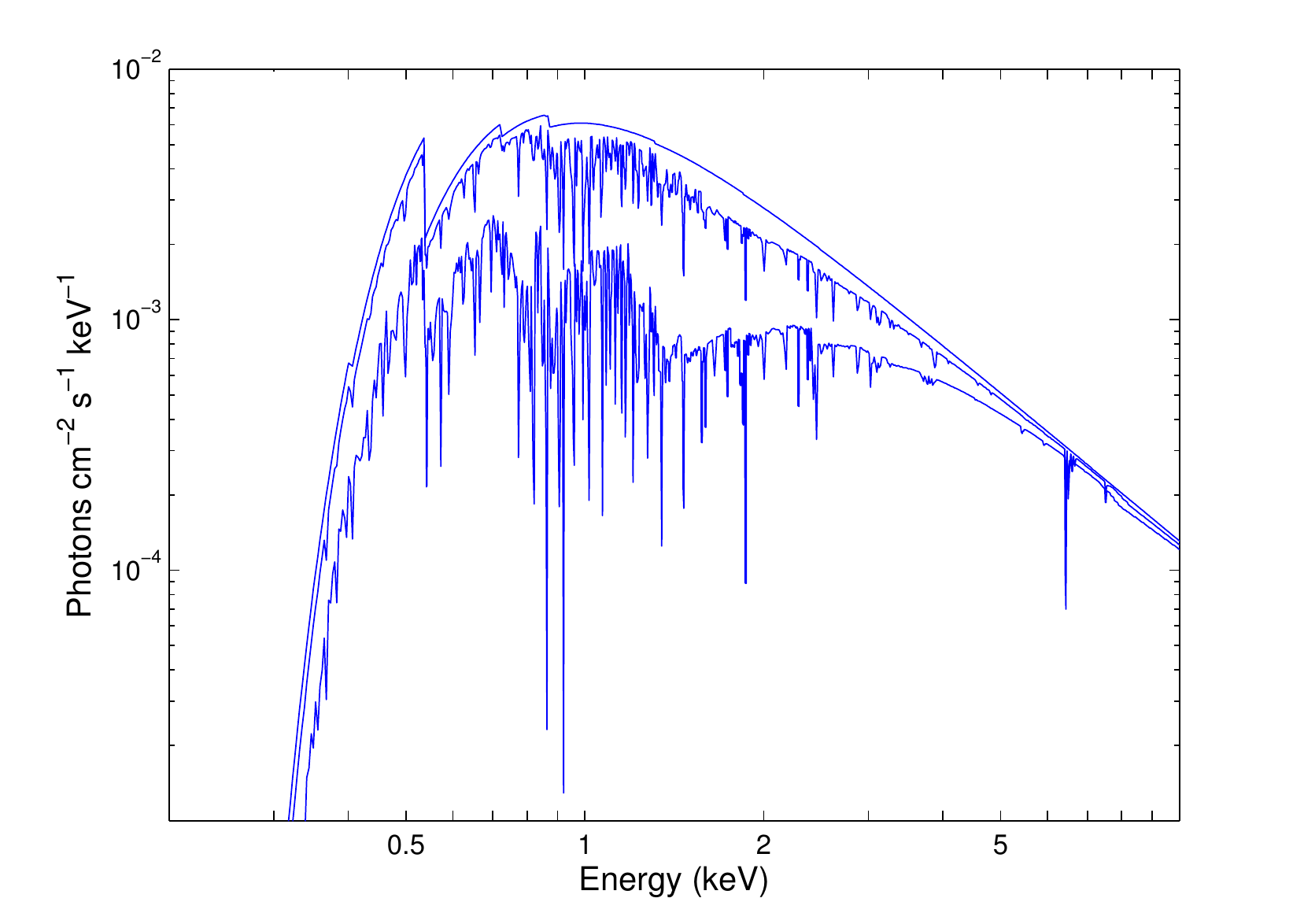}}
\caption{The warm absorber model that fit the persistent (top),
shallow-dipping (middle) and deep-dipping (bottom) spectra of \src\
(see Sect. \ref{warmabssec}). More absorption lines and edges are
present during deep dipping, when the ionization level is lower.}
\label{warmabsmodel}
\end{figure*}

As an alternative, we tested the hypothesis that the spectral changes
during dipping could be due to changes in the properties of both a
neutral and an ionized absorber. We used the photo-ionization code
CLOUDY version 07.02 \citep{ferland98pasp} to simulate ionized clouds
standing in the line of sight toward the X-ray emission. Cloudy
requires specifying the hydrogen density ($10^{8}$~cm$^{-3}$), an
abundance table \citep{anders89gca}, the hydrogen column density (the
hydrogen density integrated along the line of sight), an ionizing
continuum, and the ionization parameter defined by \citet{tarter69apj}
as

\begin{equation}
\xi = \frac{L_{ion}}{n_{H}\times r^{2}} ~erg ~s^{-1}~cm
\end{equation}

where $n_{H}$ is the hydrogen density at the illuminated face of the
ionized cloud and $r$ is the source cloud separation. $L_{ion}$ is the
luminosity between $1 ~Ryd$ and $1000 ~Ryd$.

As ionizing continuum, we assumed a power law with an X-ray slope
consistent with what we determined from the persistent emission. We
calculate a grid of warm absorbers each having a distinct column
density/ionization parameter combination. For a fixed hydrogen
density, the ionization parameter is basically a measure of the
overall strength of the ionizing continuum and consequently, it
affects the presence/absence of the different ionic species. On the
other hand, the hydrogen column density controls the amount of the
ionic species present. Together these two quantities control the
amount of absorption, at every energy, suffered by the incident
continuum upon exiting the warm absorber. The transmitted continuum
bears the spectral signatures, such as absorption edges and lines, of
the ions present in the warm absorber. XSPEC interpolates over the
transmitted continua from the warm absorber grid to find the best fit
to a spectrum.

To test the hypothesis that the spectral changes during dipping could
be due to changes in the properties of both a neutral and an ionized
absorber, we kept the parameters of the power law linked between
persistent, shallow-dipping and deep-dipping spectra, but left the
parameters of the foreground absorber and of the ionized absorber
free. However, since the persistent spectrum is consistent with a
power law and does not show any absorption features due to highly
ionized elements (see Sect. 3.3), the properties of the warm absorber,
if present during persistent emission, cannot be constrained from this
observation. Consequently, for the persistent spectrum, we fixed the
warm absorber parameters to values such that the plasma is transparent
at all energies, as if it were absent. In practise, we fixed the
ionized column density to $10^{21.5}$~cm$^{-2}$ and the ionization
parameter to $10^{5.0}$~erg~s$^{-1}$~cm, which indeed corresponds to a
thin nearly fully ionized and hence transparent medium.

Fig. \ref{fitwarmabsorber} shows the spectra and best fit model and
Fig. \ref{warmabsmodel} shows the unfolded spectra. Table~2 gives the
values of the fit parameters. The values of the photon index and the
normalization of the power-law component were 1.98 and
$1.2\times10^{-2}$ ~photons ~keV$^{-1}$ ~cm$^{-2}$ ~s$^{-1}$
respectively. They are consistent with the values obtained when
fitting the persistent spectra alone. The foreground neutral
absorption increased slightly from $(0.410 \pm 0.007)\times 10^{22}$
~cm$^{-2}$ during persistent interval to $(0.420 \pm 0.009)\times
10^{22}$ ~cm$^{-2}$ during shallow dipping and to $(0.45 \pm
0.03)\times 10^{22}$ during deep dipping, which represents, between
each two consecutive intervals, a change at a level of $1.4~\sigma$
and $1.7~\sigma$ respectively.  The ionization parameter decreases
from $10^{2.52}$~erg~s$^{-1}$~cm during the shallow-dipping interval
to $10^{2.29}$~erg~s$^{-1}$~cm during the deep-dipping emission, which
represents a change at a level of 5~$\sigma$.  The column density of
the ionized absorber increases from $4.3_{-0.5}^{+0.4} \times 10^{22}$
~cm$^{-2}$ during the shallow dipping to $11.6_{-0.6}^{+0.4} \times
10^{22}$ ~cm$^{-2}$ during the deep-dipping interval, which represents
a change at a level of more than 10~$\sigma$. These results indicate
that the spectral changes during dipping in \src\ are consistent with
an increase in column density and a decrease in ionization state of a
highly ionized absorber, associated with a slight increase in column
density of a local neutral absorber.


\subsection{Model comparison near $\sim$1 keV}
\label{interpretation}

\begin{figure}[!th]
\centerline{\includegraphics[angle=0,width=0.5\textwidth]{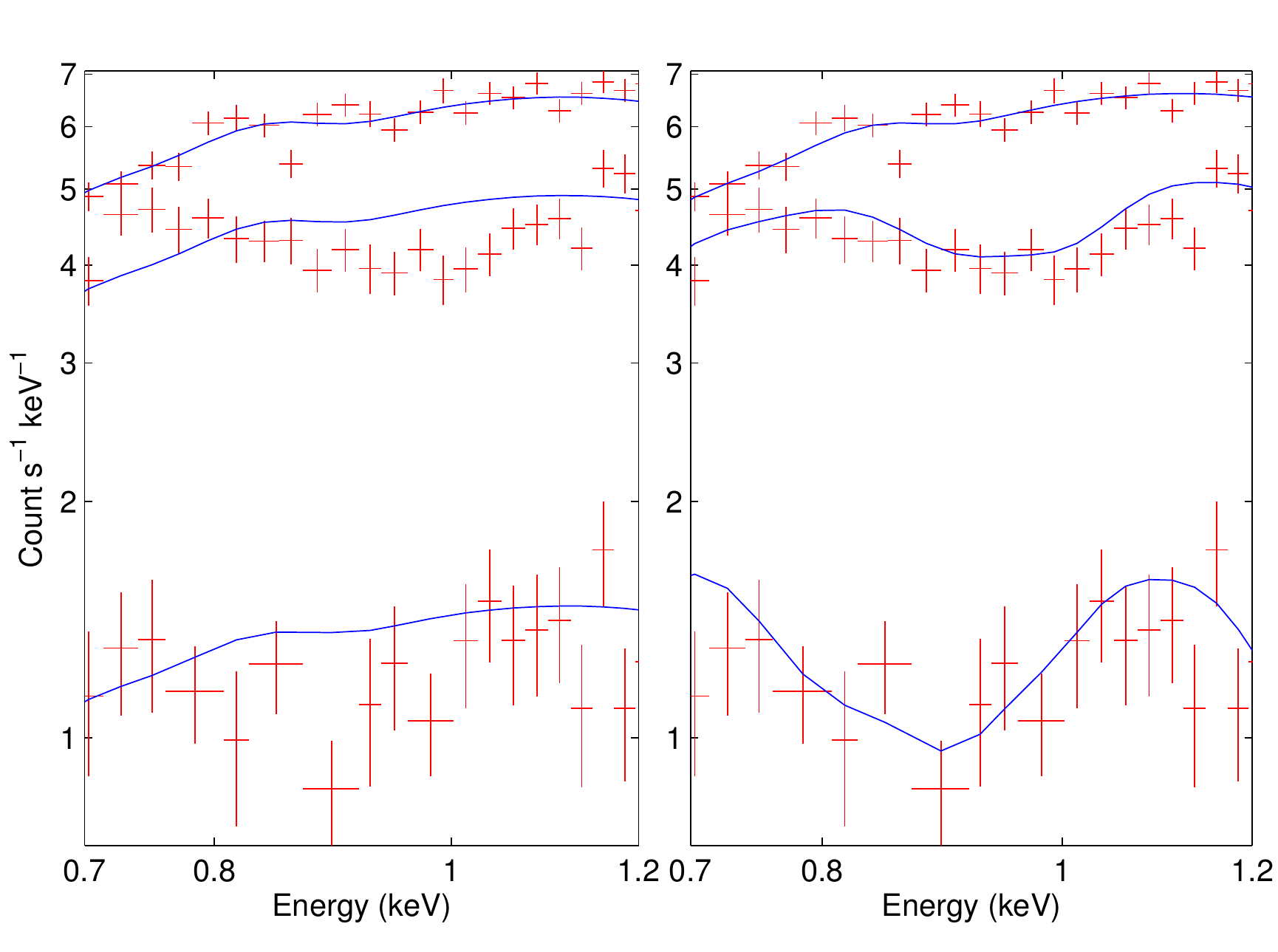}}
\caption{A zoom at the 1 keV region of the persistent, shallow-dipping and deep-dipping spectra fitted with either, the partial covering model (left panel) or the ionized absorber approach (right panel). The warm absorber model fits better the 1 keV region of the dipping spectra.}
\label{zoom1keV}
\end{figure}

Although the partial covering model (Sect. \ref{parcovsec}) and the
warm absorber model (Sect. \ref{warmabssec}) can both fit the spectra
of \src, we note that the warm absorber model gives a better fit,
especially near $\sim$1~keV. Fig. \ref{zoom1keV} shows a zoom at 1 keV of
the \src~ spectra ~fitted with the ionized absorber (right panel) and
the partial covering model (left panel). 

The modulation of the shallow and deep-dipping spectra, near 1 keV and
0.9 keV, respectively, is clearly better accounted for by the warm
absorber model. Restricting the energy range to 0.7--1.2~keV, the
reduced chi squared would be 0.99 (50 d.o.f) for the warm absorber and
1.84 (52 d.o.f.) for the partial covering model. This strongly favors
the warm absorber interpretation. Indeed a "depression" such as seen
in the dipping spectra near 1~keV is most easily explained by invoking
absorption than by a combination of emission components. Now,
absorption by cool material does not predict strong edges near 1~keV
that would be consistent with the rest of the spectral shape. On the
contrary, absorption by a warm absorber naturally includes many lines
and edges from ionized species (see Fig. \ref{warmabsmodel}).  The
convolution of these narrow features with an instrument response with
moderate energy resolution naturally translates into a depression such
as observed in the pn dipping spectra of \src.


\section{Discussion}

We presented a detailed X-ray light curve covering one orbital period
for the LMXB \src. It shows irregular dipping activity and a total
eclipse, implying that the system is viewed with a high inclination
angle. We fitted the persistent spectrum with an absorbed power law
with a photon index of $1.94\pm0.02$ while the foreground hydrogen
column density was $(0.401\pm0.007)\times 10^{22}$~cm$^{-2}$. We did
not detect absorption lines from highly ionized species such as
\fetfive\ resonant line at 6.65~keV seen in other dippers, but the
upper limits we derived on their EW are not constraining. The overall
0.2--10~keV spectrum becomes harder during the dips, but this change
is inconsistent with a simple increase in absorption by cool
material. We modeled the spectral changes between persistent and
dipping intervals in two ways.

In a first place, we modeled the spectral changes during dipping by
the partial covering of the power-law emission. The covering fraction
increased from shallow-dipping to the deep-dipping interval. However,
the neutral column density of the covering component decreased from
shallow dipping to deep dipping. The increase of the covering fraction
is consistent with the results obtained for other dippers studied with
the partial covering model in the past
\citep[\eg~][]{1624:churchaa}. However, the decrease in column density
is inconsistent with what is commonly found in LMXBs, where the column
density also increases as we go deeper in the dips
\citep[\eg~][]{XBT0748:church98apj,XB1254:smale02apj}.

In a second place, we modeled the spectral changes during dipping by
the variation in the properties of an ionized absorber and of a
neutral absorber. The foreground column density increases slightly
from $(0.410 \pm 0.007)\times 10^{22}$ ~cm$^{-2}$ during the
persistent emission to $(0.45 \pm 0.03)\times 10^{22}$ ~cm$^{-2}$
during deep dipping. The ionization parameter decreases from
$10^{2.52}$~erg~s$^{-1}$~cm during the shallow-dipping interval to
$10^{2.29}$~erg~s$^{-1}$~cm during deep dipping. On the contrary, the
column density of the ionized absorber increases from
$4.3_{-0.5}^{+0.4} \times 10^{22}$ ~cm$^{-2}$ to $11.6_{-0.6}^{+0.4}
\times 10^{22}$ ~cm$^{-2}$. The properties of the ionized and neutral
absorbers that we derived for \src, and their change during dipping,
are similar to those found in the other dippers observed with \xmm\
\citep{1323:boirin05aa,diaztrigo06aa}, especially to those of \mxb. In
this source, the ionization parameter decreases to reach
$10^{2.42}$~erg~s$^{-1}$~cm during the deepest dips while the ionized
absorber column density increases to $53\times 10^{22}$ ~cm$^{-2}$. We
note that the change in the foreground column density, $N_{H}^{fore}$,
between persistent and the deepest dips in \src\ is very small
compared to \mxb. This could partially be due to the fact that we
split the dips into two dipping levels while the dips in \mxb\ were
split into five levels. Additionally, \citet{diaztrigo06aa} were able
to explain the spectral changes in the 0.2--10 keV continuum with the
change in the properties of a warm absorber and a neutral one from
\useventeen\ and \exo, the two other sources that do not show any
spectral signatures due to highly ionized species near 7 keV.

Although the partial covering model and the warm absorber model can
both fit the spectra of \src, the depression in the dipping spectra
near 1~keV is clearly better accounted for by the warm absorber model.
It is naturally explained by the many lines and edges produced by the
ionized absorber (and not by the neutral one), but not resolved by the
EPIC--pn instrument. 

Similar depressions near 1 keV were detected in the pn spectra of
several other dippers (\nineteen, \exo, \twelve\ and \mxb) and
interpreted in the same way \citep{diaztrigo06aa}. However, the signal
to noise ratio of the RGS spectra acquired simultaneously for these
sources was often too low, especially during dipping, to enable to
confirm the interpretation by resolving the individual lines or edges
that would constitute the depression seen in pn spectra. Among the
dippers studied by \citet{diaztrigo06aa}, only \mxb\ had shown clear
absorption lines in the RGS \citep{1658:sidoli01aa}, but this was
during persistent emission, while too few counts were detected by RGS
during dipping. We note however, that, before the use of warm absorber
models, \citet{1916:boirin03aa} modeled a feature appearing in the pn
spectra of \nineteen\ with an edge at an energy decreasing from
$0.98\pm0.02$~keV to $0.87^{+0.06}_{-0.04}$~keV from persistent to
deep dipping, and interpreted it as resulting from edges of moderately
ionized Ne and/or Fe, with the average ionization state
decreasing. The RGS spectra could also be accounted for by including
an edge at an energy lower during dipping than during persistent
intervals, consistently with pn results.

The absorption feature near 1 keV in the dipping spectra of \src,
together with the fact that warm absorbers were unumbigously
identified thanks to the detection of some of their narrow spectral
signatures in many dipping sources whose spectral changes used to be
modeled by partial covering, strongly favor the warm absorber
approach. This interpretation would be easily confirmed by a longer
XMM observation of \src\ that would enable the dectection of, \eg, the
\fetfive\ absorption line at 6.65~keV down to an EW of about 15~eV.


\bibliographystyle{aa}

\end{document}
